\documentclass{ws-mplb}

\begin{document}

\markboth{Victor Lakhno}{Peculiarities in the concentration dependence
of the superconducting transition temperature ...}

%
\catchline{}{}{}{}{}
%

\title{PECULIARITIES IN THE CONCENTRATION DEPENDENCE OF THE SUPERCONDUCTING
TRANSITION TEMPERATURE IN THE BIPOLARON THEORY OF COOPER PAIRS}

\author{VICTOR LAKHNO}

\address{Institute of Mathematical Problems of Biology RAS \\
the Branch of Keldysh Institute of Applied Mathematics of Russian Academy of Sciences\\
142290, Vitkevicha str. 1,  Pushchino, Moscow Region, Russia\\
lak@impb.psn.ru}

\maketitle


\begin{abstract}
It is shown that the bipolaron theory of Cooper pairs suggests that there
is a possibility for a superconducting phase to exist at low and high levels
of doping and be absent at intermediate level of doping. The results
obtained imply possibly universal character of 1/8 anomaly.
\end{abstract}


The results of paper,\cite{1} where a Cooper pair was demonstrated to be nothing but a bipolaron, actualize the problem of a bipolaron in a polaron gas.  There consideration was given to a problem of electron-phonon interaction (EPI) between two electrons in Cooper's formulation\cite{2} when the Fermi energy exceeds the EPI one: $E_F>\left|E_{pol}\right|$, where  $E_{pol}$  is a polaron energy. In high-temperature superconductors (HTSC), however, of importance is the case of $E_F<\left|E_{pol}\right|$  (HTSC with low level of doping). We will show that these two cases lead to two qualitatively different pictures.

a) The case of  $E_F>\left|E_{pol}\right|$

Let us consider the limit case when $E_F>>\left|E_{pol}\right|$  and above the Fermi surface there is one electron taking part in  electron-phonon interaction. In view of Pauli principle the interaction of this electron with the electron occuring below the Fermi surface can be neglected. Hence, in this case we have the polaron problem for an electron occuring near the Fermi surface. Because of EPI, the energy of this electron should be below the Fermi surface at the depth of $E_{pol}$. But the same will be valid for all the electrons occuring at the Fermi level: owing to EPI their energy will be decreased by $E_{pol}$. Hence, if we denote the Fermi energy in the absence of EPI by  $E_{F}^0$, then in the presence of EPI the renormalized value of the Fermi energy will be:
 $E_F=E^0_F+E_{pol}$. The masses of electrons whose energies occur near $E_F$  will also undergo a relevant polaron renormalization. Therefore in the energy layer  ($E_F+E_{pol}$, $E_F$) we will have a polaron gas.

Let us consider the case of two electrons above the Fermi surface. From the aforesaid it follows that now the Fermi surface is determined by $E_F$ rather than by  $E^0_F$. Now the presence of EPI cannot decrease the energy of either of the two electrons by the value of  $E_{pol}$ since in view of EPI, the energy of the electrons occuring on the Fermi surface is already decreased by this value. If the two electrons form a paired state, the energy of the state will be below the Fermi surface $E_F$  at the depth of  $E_{bp}$, where  $E_{bp}$  is the bipolaron energy for any value of the EPI constant   $\alpha$. This result is in complete agreement with Cooper's conclusion\cite{2} about instability of the  Fermi surface with respect to formation of pairs for arbitrarily small values of  $\alpha$, which makes the polaron theory in metals qualitatively different from that in polar dielectrics. Accordingly, the value of the gap near the Fermi surface will be not $\left|E_{bp}-2E_{pol}\right|$  as it is resume there,\cite{1} but   $\left|E_{bp}\right|$. Hence, only the electrons whose energies occur in the ($E_F+E_{pol}$, $E_F$) layer take part in the formation of bipolaron paired states, which differentiates the bipolaron theory of superconductivity\cite{3} from the BCS theory\cite{4} which implies that for  $T=0$ all the electrons are in the paired state.

b) The case of   $E_F<\left|E_{pol}\right|$

In this case we cannot neglect the interaction between the excess electrons and electrons below the Fermi surface. If , as in the previous case, we believe that the electrons below the Fermi surface are polarons then the Cooper problem will correspond to the problem of a bipolaron in a polaron gas. As a bipolaron is placed in a polaron gas, an additional energy difference arises in view of the fact that a bipolaron is a Bose particle while a polaron is a Fermi particle.

Hence if we take $\Delta E=E_{bp}-2E_{pol}$ (corresponding to the energy gain for   $\Delta E<0$ or energy failure for  $\Delta E>0$  in the absence of a polaron gas) as a reference point of energy in the absence of polaron gas, then in a polaron gas  $E_{pol}$  should include the additional terms  $E_F=p^2_F/2m_{pol}$ ( $p_F=\left(3\pi^2\right)^{1/3}\hbar n^{1/3}$ is the Fermi momentum, $n$ is the concentration of current carriers) and  $E_{exch}$, where  $E_{exch}=-e^2p_F/\pi\hbar\epsilon$  is the exchange energy in the Hartree-Fock approximation,\cite{5}  $m_{pol}$ is the polaron mass,  $\epsilon$ is the dielectric permittivity. As a result, the bipolaron stability criterion takes on the form:\cite{6,7} $\Delta E<2E_F+2E_{exch}$.

This implies that on condition that: $\left(m_{pol}e^2/\pi\hbar\epsilon\right)^2>-m_{pol}\Delta E>0$ the bipolarons are stable in two regions:$\left(0,p_{F1}\right)$  and $\left(p_{F2},\infty\right)$, $p_{F1,2}=m_{pol}e^2/\pi\hbar\epsilon\pm\sqrt{\left(m_{pol}e^2/\pi\hbar\epsilon\right)^2+m_{pol}\Delta E}$, where: $p_{F1}$  corresponds to the sign ($-$), and  $p_{F2}$ - to the sign ($+$).
The region ($0$, $p_{F1}$) corresponds to small concentration of current carriers (low doping) while the region ($p_{F2}$, $\infty$ ) corresponds to large concentration (high doping). If we believe that the presence of bipolarons at $T=0$  immediately leads to superconductivity, then the existence of two different regions of bipolaron stability will correspond to the existence of two different regions of superconductivity occurrence.

If the converse condition is fulfilled: $\left(m_{pol}e^2/\pi\hbar\epsilon\right)^2<-m_{pol}\Delta E<0$ the bipolarons are stable at any level of doping. In this case the concentration dependence of the critical temperature of the superconducting transition $T_C$  will have a local minimum.

This dependence is actually realized in superconductors La$_{2-x}$M$_x$CuO$_4$, $M=\left(Sr,Ba\right)$ , where the high-temperature superconductivity was observed for the first time. For example, in La$_{2-x}$Sr$_x$CuO$_4$ the optimal level of doping is equal to $x\approx0,16$. As $x$ decreases, $T_C$  is lowered too. This behavior remains unchanged up to $x\approx1/8$ , when  $T_C$  reaches its minimum. As $x$ further decreases,  $T_C$ grows achieving some maximum and then decrease vanishing at small $x$. In La$_{2-x}$Ba$_x$CuO$_4$  this behavior is still more pronounced: there exists a sharp dip in the $T_c-x$ phase diagram, indicating that bulk superconductivity is greatly suppressed in narrow range around $x=1/8$.

The emergence of minimum in the concentration dependence is known as $1/8$ anomaly which probably has universal character being observed in other HTSC materials.\cite{8,9,10,11}

\section*{Acknowledgements}
The work was supported by RFBR, N 16-07-00305 and RSF, N 16-11-10163.


\begin{thebibliography}{0}
\bibitem{1} V. D. Lakhno, \textit{Mod. Phys. Lett. B} \textbf{30} (2016) 1650365.
\bibitem{2} L. N. Cooper, \textit{Phys. Rev.} \textbf{104} (1956) 1189.
\bibitem{3} V. D. Lakhno, \textit{SpringerPlus} \textbf{5} (2016) 1277; V. D. Lakhno, arXiv:1510.04527 [cond-mat.supr-con].
\bibitem{4} J. Bardeen, L. N. Cooper, J. R. Schrieffer, \textit{Phys. Rev.} \textbf{108} (1957) 1175.
\bibitem{5} N. W. Ashcroft, N. D. Mermin, \textit{Solid State Physics} (Holt, Rinehart and Winston, New York, 1976).
\bibitem{6} A. A. Shanenko et al., \textit{Sol. St. Comm.} \textbf{98} (1996) 1091.
\bibitem{7} M. A. Smondyrev et al., \textit{Phys. Rev. B} \textbf{63} (2000) 024302.
\bibitem{8} A. R. Moodenbaugh et al., \textit{Phys. Rev. B} \textbf{38} (1988) 4596.
\bibitem{9} S. A. Kivelson et al., \textit{Rev. Mod. Phys.} \textbf{75} (2003) 1201.
\bibitem{10} M. Vojta, \textit{Adv. Phys.} \textbf{58} (2009) 699.
\bibitem{11} M. H\"{u}cker et al., \textit{Phys. Rev. B} \textbf{83} (2011) 104506.
\end{thebibliography}
\end{document}